\documentclass[a4paper, 12pt, titlepage]{article}
\usepackage[dvips]{graphicx}

\def\supit#1{\raisebox{0.8ex}{\small\it #1}\hspace{0.05em}}
\usepackage[a4paper, left=2cm, right=2cm]{geometry}

\title{Development of a carbon fibre composite active mirror:
Design and testing}
\author{S. Kendrew\supit{a}\footnote{Email: sk@star.ucl.ac.uk, Fax: +44 207 679 7153}, P. Doel\supit{a}, D. Brooks\supit{a}, C. Dorn\supit{b}, C. Yates\supit{b}, R.M.
Dwan\supit{b},\\I. Richardson\supit{c}, G. Evans\supit{c}\\[3pt]
\small\supit{a}University College London, Department of Physics
and Astronomy,\\
\small Gower Street, London WC1E 6BT, United Kingdom\\
\small\supit{b}QinetiQ, Cody Technology Park, Ively Road,\\
\small Farnborough, Hants GU14 0LX, United Kingdom\\
\small\supit{c}Cobham Composites, Gelders Hall Road,\\
\small Shepshed, Leics LE12 9NH, United Kingdom}
\date{}

\begin{document}
\maketitle \abstract {Carbon fibre composite technology for
lightweight mirrors is gaining increasing interest in the space-
and ground-based astronomical communities for its low weight, ease
of manufacturing, excellent thermal qualities and robustness. We
present here first results of a project to design and produce a 27
cm diameter deformable carbon fibre composite mirror. The aim was
to produce a high surface form accuracy as well as low surface
roughness. As part of this programme, a passive mirror was
developed to investigate stability and coating issues. Results
from the manufacturing and polishing process are reported here. We
also present results of a mechanical and thermal finite element
analysis, as well as early experimental findings of the deformable
mirror. Possible applications and future work are discussed.}

\setlength{\parskip}{3ex}
\section{Introduction}
The last decade has seen a substantial effort to develop
ultra-lightweight optics for imaging systems. Research has been
strongly driven by the aerospace industry, where the weight of the
payload directly influences the cost of a mission or craft. The
advent of the James Webb Space Telescope (JWST) in particular has
invigorated the research drive \cite{stahl}. In astronomy too,
where instruments and telescopes have dramatically increased in
size and complexity over this period, using lightweight optics
allows decreased tolerances for support structures, which can
again save substantial cost as well as reduce maintenance
requirements for the telescope.
\par
Several methods exist for reducing mass of traditional optical
materials, such as Zerodur and Beryllium, by material removal
\cite{zerodur1}.  Lightweighting conventional low thermal
expansion glass mirrors is however expensive and there is
considerable variability in the time taken to produce and polish
glass items, and limitations of material suppliers - all of which
increase programme risks. Beryllium is also toxic and requires
very strict safety precautions during the manufacturing process.
Much progress has been made in recent years in developing novel
materials for lightweight optics, both for space and terrestrial
environments - including carbon fibre-reinforced silicon carbide
(CeSic) \cite{cesic} and sintered silicon carbide
(SSiC)\cite{ssic}. \par

Carbon fibre composite (CFC) materials offer an attractive
alternative, being very robust whilst relatively easy (and
cost-effective) to manipulate. Because of their composite nature
the material properties can be tailored to a specific application,
thus making them very versatile. They also exhibit near-zero
thermal expansion. Table 1 shows a comparison of typical material
properties for CFC and other mirror materials. In general,
composite structures are manufactured using a moulding process.
Plies of fibres in a raw resin material are deposited onto a mould
or mandrel together with a curing agent. Heat and pressure are
then applied, eventually resulting in a 'quasi-isotropic'
laminate.

However, for use in the optical wavelength regime, several
problems still remain to be overcome, one of which is the surface
roughness. Carbon fibre composites can not be polished directly.
To achieve a good surface roughness, replication techniques must
be used using high optical quality moulds or the composite must be
coated with a polishable material. Mirror form error presents
another problem. Because the curing process takes place at
elevated temperature, there are issues with the overall form on
release from the mould.

\par
Several research projects into carbon fibre composite mirrors
(both active and passive) are ongoing
\cite{cfcmirrors1,cfcmirrors2}, particularly in the United States,
focused largely on producing ultra-smooth surfaces using
replication techniques. This work is important, as ease of
replication can significantly speed up the production process when
a large number of optics is required, as well as cut costs. For a
high-quality optical system, however, a good form is also
required. If actuators are used to control the mirror's shape,
tolerances can be decreased as low-order shape distortions can be
compensated - but if part of the actuators' stroke is required to
flatten the mirror this will decrease their ability to compensate
for other optical effects.

An alternative approach is to coat the mirror with a material that
can be ground and polished. The advantage of this is that the
tolerance on the production of the CFC mirror from the mould are
reduced as the substrate can be ground to the required form either
before or, if sufficient coating material is applied, after the
coating process (see, for example, \cite{abusafieh}). This paper
will describe in detail the first results from a project carried
out at University College London (UCL) in collaboration with
industrial partners QinetiQ and Cobham Composites to develop a
prototype deformable CFC optical mirror using this approach. The
aim of this project was to investigate the production methods and
to test key performance parameters, such as mirror influence
functions and form errors. The following sections will describe
the results from a passive test mirror, design of the prototype
active mirror, potential applications and future work.\par\bigskip

\section{Design considerations}

The ultimate aim of a deformable mirror for active or adaptive
optics is an ability to correct any aberrations originating in the
atmosphere or within the optical system itself. These aberrations
can arise from changes in the gravitational or thermal environment
in the case of a space-borne system, or from gravity sagging or
atmospheric turbulence for ground-based systems. For the CFC
active mirror, the design was driven by several considerations.
The baseline specifications (Young's modulus, stiffness, actuator
spacing) were made to match that of a previous deformable mirror
system that was developed at UCL in the late 1990s. This system is
described in detail by Lee et al. \cite{ucl_asm} and Lee \cite{
junho_thesis} and features an aluminium faceplate with a diameter
of 27 cm, whose shape is controlled by 7 magnetostrictive
actuators spaced at 10 cm in a hexagonal arrangement, plus full
backing structure, drive electronics and software. Its form is
concave and spherical, with a radius of curvature of 2945 mm. The
mechanical design of the aluminium mirror resulted from a study by
Bigelow et al. \cite{asm_study}. In addition, the CFC material
dictated its own particular requirements.\par

The faceplate's stiffness must be such that the inter-actuator
sagging during both horizontal and vertical pointing is small. For
a typical optical surface this is of the order of nanometres.
Also, the forces executed on the structure by the actuators must
lie within the interlaminar strength to avoid cracking and
delamination. \par Because CFC is intrinsically non-reflective, a
nickel coating was added to the design, and this formed the
biggest challenge to the project. Though the composite has an
excellent thermal stability, the addition of a nickel layer was
likely to introduce a mismatch effect with possible mirror
stresses and distortions as a result. Finite element analysis
(FEA) results indeed supported this prediction. The thickness of
the nickel layer also was to be balanced against the requirements
for polishing and grinding in the post-production phase.\par

Another important question mark hangs over the mirror's stability
through varying humidity regimes. As shown in table 1, CFC has a
non-zero moisture expansion coefficient. The exact value depends
on the matrix material used, e.g. a normal space-qualified epoxy
has a CME of around 25 parts per million (ppm) (for a fibre volume
fraction of 0.6) whereas a cyanate ester resin's is just 9-10 ppm,
albeit at 3 times the cost \cite{ian}. Little information is
available on how the material's moisture absorption affects its
strength or shape; as with many aspects of carbon fibre materials,
the effect is poorly quantified. This was hence not a major
influence on the proposed design but the mirror's stability was
closely watched during testing. Care was taken to avoid moisture
uptake by the substrate during polishing by sealing the mirror
edge with a lacquer coating.\par To investigate some of these
issues and test the manufacturing and polishing methods, it was
decided to produce a passive composite test mirror with the same
size and radius of curvature. The actual laminate configuration
for this mirror was stiffer than the active design as its shape
was not intended to be actively controlled. Such a mirror could be
of use in space applications where active control is beyond the
mission's scope.

\section{Passive test mirror: Design and testing}
\subsection{Preliminary FEA results}\label{FEA}
For the passive mirror, a sandwich plate design was decided on
from the outset to make optimal use of the composite's low weight
and thermal expansion properties. Commonly used in composite
manufacturing, a cored lay-up has the advantage of providing extra
stiffness to the laminate without adding significantly to its
weight. For this particular purpose, an aluminium alloy honeycomb
was chosen, as these are cheap, widely available, easy to
manipulate, and have a very good strength-to-weight ratio.\par

Finite element analysis was then carried out to predict the amount
of gravitational sagging to be expected from laminates with
different core thicknesses. Initial modelling was carried out at
QinetiQ using Ansys/Nastran, later reconfirmed by further work at
UCL presented here, using I-Deas. The mirror itself was modelled
using 2D thin shell elements. The laminate was constructed using
the Laminates tool in I-deas, which uses the material properties
of individual plies to compute equivalent orthotropic properties
for the laminate. Using this method the laminate thickness is
automatically translated to the 2D elements. Preliminary tests
showed that individual ply thickness or stacking sequence -
providing the ply balance and symmetry was respected - did not
have a significant effect on resultant material properties.

Three different support methods were modelled: an edge support
round the entire diameter, a 3-point support at the mirror edge,
and a 7-point support. In the 7-point support model, nodes within
a circular area of diameter 25 mm were coupled and restrained at
the proposed actuator locations, to simulate the effect of an
actuator pad. The reason for including this configuration was to
enable an easy comparison with earlier FEA results for the
aluminium active mirror, upon which the design was based. For a
passive mirror an edge or 3-point support would be more realistic.
The approximate restraint method was used to save on computing
time; to include a full model of actuators attachments would have
increased the solve time significantly. Table \ref{passive_cores}
summarizes the results for rms gravitational sag for various core
thicknesses. Figure \ref{fea_pic} shows a plot of gravitational
sag results from FEA for the 10 mm cored model with 7-point
support. A similar FEA model of the aluminium active mirror
\cite{junho_thesis} showed a maximum zenith-pointing sag of 31.9
nm. The 10 mm cored model matched this value very closely with a
maximum sag of 33.4 nm and this lay-up was therefore chosen for
the test mirror.
\par

Thermal FEA on the 10 mm cored model yielded a predicted defocus
of 16-17 mm for a $100 \mu m$ nickel layer for $\Delta T=+10C$,
reducing to just 5 mm if the layer is reduced to $25 \mu m$.
Models also showed that a matching nickel layer on the mirror's
back surface would virtually eliminate the effects of this
mismatch. However, the manufacturing modifications required for
this were not part of the project's aims.\par

\subsection{Design and initial form}
During the manufacturing phase carried out by Cobham Composites a
nickel coating was transferred onto the mirror from the mould. A
picture of the mirror prior to grinding and polishing is shown in
figure \ref{passive}. The coating thickness was overspecified in
the first instance to ensure that enough material was available
for grinding and polishing; experience with the test mirror could
indicate whether it could be reduced to minimise any bimetallic
effects.
\par

Before grinding and polishing the most marked feature on the
mirror's surface was a fibre print-through. The origin of this was
uncertain as print-through was not expected through a 100 $\mu m$
thick layer of metal. Following discussion with the manufacturing
team it was suggested that the features were due to post-cure
shrinkage of the adhesive layer between composite and nickel that
'pulled' the nickel around the underlying fibres. Also present
were uni-directional striations across the reflective surface
spaced around 1 cm apart, and a ring-shaped depression
approximately 2 cm inside the mirror edge. Figure
\ref{profilometer} shows a profilometry measurement with a Mark 1
Form Talysurf before grinding and polishing. This measurement
shows form errors of approximately $\pm 5 \mu m$ from a spherical
form. The ring feature can be seen as the first valley in the
plot. After an investigation of the materials used it appeared
that a filler material had been used along the edges to protect
the aluminium core from crushing under the vacuum bag pressure
during the lay-up process. The cause of the striations was not
conclusively identified.

Interferometry testing revealed a radius of curvature of 2910 mm -
35 mm shorter than that of the mould - indicating that residual
stresses on release from the mould had increased the mirror's
concavity. This can almost certainly be attributed to stresses
caused by the thermal expansion mismatch between the nickel and
composite on cooling. Indeed, a composite-only test sample made
with the same mould did not show this level of radius of curvature
shrinkage.
\par

\subsection{Final surface, form and testing}

The mirror was ground and polished manually by D. Brooks using
polishing tools of up to $7/8$ of the mirror size. The edges were
protected against moisture ingress from the polishing slurry. The
mirror was supported on a compliant layer whilst being polished to
prevent the introduction of further stresses. After successive
cycles of grinding, polishing and measurement, a surface roughness
of 4 nm was obtained, measured over a $1 mm^2$ square. This value
is no absolute limit and can be improved on with experience. The
striations and fibre print-through observed after the lay-up were
removed in the polishing process.

Figure \ref{passive_postform} shows the final mirror form. The
interferogram clearly shows a residual edge depression that was
not removed by the grinding process. The overall surface form was
$1 \mu m$ P-V, the edge ring accounting for a large part of this.
The central 24 cm displayed a form of $517 nm$ P-V. Apart from the
bimetallic effect, another possible cause of these form errors is
a slight misalignment of ply angle within the composite laminate,
which could affect its stability. Figure \ref{passive_final} shows
a picture of the passive mirror after polishing.

The mirror displayed no measureable form changes over one month at
constant temperature and relative humidity. Its long term
stability remains under investigation. A thermal stability test
showed a defocus of 6 mm for $\Delta T = 5C$. The magnitude of
this is in good agreement with FEA results.\par

\section{Active mirror design}

From experience with the passive mirror, it was clear that the
main problem with a nickel-coated active mirror was likely to be
the thermal stability. Flexures were included in the design to
relieve any stresses. The amount of actuator stroke used to
flatten the mirror should be kept to a minimum so efforts must be
made to obtain the best possible form before flattening.\par

The key property that defines how a material will respond to
actuation (in terms of stroke and influence function shape) is its
specific stiffness, defined as $E/\rho$, with $E$ the Young's
modulus and $\rho$ the material density. A higher specific
stiffness increases the actuator coupling and avoids a 'dimpling'
effect of the faceplate, but also increases the required power for
a given stroke, and hence also the dissipation, which can be a
critical factor.\par

Because of the relatively large actuator spacing of 10 cm and to
counter the bimetallic effect, the composite was stiffened with an
aluminium honeycomb core, as with the passive test mirror. In this
case, however, the core thickness was just 4 mm and no core filler
was used to avoid the edge effects encountered previously. Similar
FEA models were carried out to those describe in section
\ref{FEA}. For the chosen laminate these revealed an rms
gravitational sag of 150 nm with the mirror pointed in the
vertical direction and the actuators modelled as points, as
before. The mirror was manufactured by Cobham Composites and
delivered to UCL late in 2004, for grinding, polishing and
testing. On delivery it weighed 312 g (see figure
\ref{active}).\par

Initial interferometry testing showed that the mirror showed a
degree of astigmatism ($65 \mu m$) and again some print-through
from the fibres. Grinding and polishing reduced the former to $15
\mu m$ and eliminated the fibre print-through. No edge ring was
observed, suggesting that the team's assessment of the feature in
the passive mirror was accurate. At the time of writing the mirror
was being prepared for active testing. The magnetostrictive
actuators will be connected to the mirror faceplate via circular
pads of 25 mm diameter to reduce any 'top-hat' effect in the
influence functions. A low-shrinkage (less than 1\%) glue suitable
for use with composite materials was selected for this purpose.
The aluminium backing structure provides extra stiffness and a
reference for the actuators. Figure \ref{active_diagram} shows a
schematic of the active mirror setup. Further stability testing
will also be carried out.\par

\section{Conclusions}

Experience with carbon fibre composite optical elements remains
limited, and many problems will need addressing before the
technology can become routinely implemented. The work described
here has answered many questions and highlighted some of the major
issues.\par

The passive prototype has shown that a spherical surface can be
produced with form errors of the order of $10 \mu m$ for a 'thick'
CFC mirror design. These errors are suspected to arise from
residual stresses at the nickel-CFC interface or misalignment of
plies within the laminates. However, grinding and polishing
reduced the form errors substantially. A very good surface quality
(4 nm Ra) was achieved with conventional manual polishing, and
this can certainly be improved with experience. The fibre
print-through that was initially seen on the nickel surface was
also removed. The laminate was shown to be very robust, with
minimal delamination around the mirror's edge following
substantial working. Thermal tests showed a distortion of the
surface with changing temperature, particularly in the form of
defocus as predicted by FEA modelling, but this instability could
be resolved by coating the mirror's back surface with a matching
nickel layer.\par

Testing of the active mirror shows some residual astigmatism after
grinding and polishing. The form error, however, remains well
within the actuator stroke so should not affect the ability the
flatten the mirror.\par

The cause of the residual form errors observed in both these
mirrors will have to be investigated and the manufacturing,
grinding and polishing processes refined accordingly. However, the
initial results are promising nonetheless. In the months to come,
the deformable mirror will be tested for its active performance
and these results will be discussed in detail in further
publications.\par

\section{Acknowledgements}

This work was funded by the UK's Particle Physics and Astronomy
Research Council (PPARC) under the PPARC Industrial Programme
Support Scheme (PIPSS) and by the Joint Grant Scheme (JGS). The
authors are also grateful to the Smart Optics Faraday Partnership
for their continued support of the project.

\section*{Author biographies}

Sarah Kendrew has been studying at University College London (UCL)
since 1997, gaining an MSci. in Astronomy in 2001. She is
currently studying for a PhD on lightweight deformable mirrors
with Dr. Peter Doel as a member of the Optical Science
Laboratory.\par

Peter Doel has been a lecturer at University College London since
1998, where he is involved in research in adaptive optics systems
and astronomical instrumentation. Prior to UCL he was a member of
the Durham Astronomical Instrumentation Group as a postdoctoral
research assistant from 1990 to 1998. He obtained his PhD at
Durham University in 1990.

David Brooks joined the Optical Science Laboratory at UCL in 1985,
gaining a PhD in 2001. He is responsible for optical
manufacturing, metrology and testing in the group.

Chris Dorn is a general and systems engineer with 15 years of
experience in the space industry. He has worked on several
missions from design to operations, including Earth observation
and spaceborne astronomy. Currently the leader of the Specialist
Missions Team in QinetiQ Space Division.

Chris Yates has spent over 18 years in finite element modelling
for real systems. Initially involved with modelling of
transportable bridges for Army bridging systems, he has worked on
space-related projects at QinetiQ for the last 8 years. His
analysis has addressed both macro-scale issues associated with
large solar arrays and the micro-scale issues of optical systems.

Richard Dwan is a Theoretical Physics graduate from Durham
University and during his time there he had strong links with the
adaptive optics group based at Durham. In 2002 he joined QinetiQ's
space department and has been involved with a wide range of
reflective optical and infrared systems.

Ian Richardson joined Cobham Composites in 1994. He is the Sales
Director for this company and the wider Chelton Composites Group.
Previously he has held various research positions covering the
development of high strain and high modulus PAN based carbon
fibres and their interaction with toughened epoxy matrices. He has
over 25 years experience in composite materials.

Glynn Evans is a mechanical engineer with more than 20 years'
experience in the aerospace/composite materials industry. Since
1997 he has been the Engineering Manager at Cobham Composites.

\begin{table}[h]
\centering
\begin{tabular}[h]{|r|p{12ex}|p{12ex}|p{12ex}|p{12ex}|p{12ex}|}
\hline Material & Density ($kg/m^3$) & Young's
modulus (GPa)& Specific stiffness (N.m/kg) & CTE (ppm/C) & CME @ 35\%RH\\[0.5ex] \hline
Aluminium & 2710 & 71 & 26.2 & 23.9 & 0\\
Zerodur & 2520 & 92.9 & 36.9 & 0.05 & 0\\
SiC & 3140 & 420 & 133.8 & 2.2 & 0\\
Cesic & 2650 & 220 & 83 & 2.0 & 0\\
CFC & 1170 & 101 &  86.3 & 0.2 & 9.0-25*\\
Beryllium & 1850 & 300 & 162.2 & 11.5\\
\hline
\end{tabular}
\parbox{12cm}{\caption{\footnotesize{Typical key material properties. CME:
coefficient of moisture absorption. * CME of CFC material is
strongly matrix-dependent \cite{beryllium, ian, euro50_book} }}}
\label{materials}
\end{table}

\begin{table}[h]
\centering
\begin{tabular}[h]{|c|c|c|c|}
\hline Core thickness & \multicolumn{3}{c|}{Rms gravitational sag}\\ \hline & 3-point support & 7-point support& edge support\\
\hline 5 mm & 895.5 nm & 34.3 nm & 75.9 nm \\
10 mm & 522.7 nm & 20.6 nm & 43.3 nm\\
15 mm & 378.3 nm & 15.1 nm & 31.2 nm\\
\hline
\end{tabular}
\parbox{12cm}{\caption{\footnotesize{FEA
results for rms gravitational sag of the passive carbon fibre
composite test mirror for various core thicknesses. The 7-point
support used 25 mm diameter pad-shaped restraints to simulate the
actuation pads.}}\label{passive_cores}}
\end{table}\label{passive_cores}

\begin{figure}[h]
\centering
\parbox{12cm}{\caption{Gravity sag FEA results for the passive
mirror with a 10 mm core and 7-point support as described in
section 3.1. The central pad restraint is slightly oversized, this
is due to the layout of the nodes in that part of the mesh. This
did not significantly affect the rms of maximum sag
values.}\label{fea_pic}}
\parbox{12cm}{\caption{Passive test mirror during the grinding
process.}\label{passive}}
\parbox{12cm}{\caption{Profilometry measurement of outer 10 cm
of the mirror face, showing the structure of depressed ring around
30 mm inside the edge with a magnitude of $10 \mu m$.}
\label{profilometer}}
\parbox{12cm}{\caption{Interferometry measurement of the passive test
mirror form after grinding and
polishing.}\label{passive_postform}}
\parbox{12cm}{\caption{Picture of the passive mirror after polishing.}\label{passive_final}}

\parbox{12cm}{\caption{Active mirror pre-grinding}\label{active}}
\parbox{12cm}{\caption{Active mirror setup}\label{active_diagram}}
\end{figure}

\bibliography{SKbiblio}
\end{document}